# Wideband Low-Scattering Dual-Polarized Phased Array with Stepped Ground

Yu Luo, Shi-Gang Fang, Peng-Fa Li, Yuhao Feng, Shi-Wei Qu and Shiwen Yang

*Abstract*—This communication proposes a wideband dual-polarized phased array with ultra-wideband scattering cross section (SCS) reduction. The antenna elements are loaded on a bilateral stepped ground. This ground is carefully designed in terms of height difference, step number, and length to achieve phase cancellation near the normal direction. Wideband dipoles with vertical electric coupling are designed. The radiation frequency band covers the X-band (40%) under VSWR < 2.8. Array patterns are synthesized with the two subarrays, covering the -45° ~ +45° scanning range. The monostatic SCSs of the proposed 17 × 8 array prototype have been reduced within 3.6 ~ 30 GHz, with an averaged reduction of over 19.4/18.9 dB and an averaged in-band reduction of over 15.4/16.6 dB, under the normal *x/y* polarized incident waves respectively.

## I. Introduction

In recent years, the low-scattering design of phased arrays has drawn increasing attention, since phased arrays often constitute one of the major sources of the scattering cross section (SCS) for stealth targets. In multifunctional radar systems, dual-polarized arrays are widely employed due to their capability to simultaneously transmit and receive horizontally and vertically polarized electromagnetic waves. This feature reduces the number of antenna elements and saves space, making them adaptable to complex scenarios. A low-scattering design for such kind of arrays would meet the demand for their application in stealth platforms.

Methods for SCS reduction (SCSR) can generally be classified into two categories: absorption and phase cancellation. In absorption schemes, resistive components dissipate a substantial amount of the incident power, leading to a significant reduction in the scattered power. However, this often results in a decrease in radiation efficiency [1], [2]. In phase cancellation schemes, the incident energy is not absorbed but scattered into other directions, thus maintaining an acceptable radiation efficiency. The reflection phases of the aperture are controlled to cancel out backward-scattered fields.

There are various approaches to achieve this phase control: special matching layers, such as artificial magnetic conductors (AMC), polarization-rotating surfaces (PRS), and other metasurfaces [3]-[6]; special feeding techniques, such as phase delay line optimization and impedance network optimization [7], [8]; special antenna structures, such as phase-inverting elements and chessboard arrays [9], [10]; special covers, which adjust the refractive index distribution of the material through methods like quasi-conformal transformation optics (QCTO) [11], [12] with discrete stepped structure; special grounds, whose low-scattering characteristics are realized through the mutual cancellation of the fields scattered by different parts of the ground in the normal direction. This mechanism ensures that SCSR is independent of incident polarization, which is advantageous for low-scattering dual-polarized arrays, surpassing many SCSR methods which are functional only for one polarization.

On the one hand, conformal arrays exhibit inherent low-scattering characteristics due to their ability to conform to arbitrary surfaces, which introduces height variations, such as curved surfaces with large curvature. In [13] and [14], large bending angles provide more height difference, which contributes to a lower SCS. However, the radiating elements are only partially excited to avoid appearance of large side lobes. In [14], only three rows of elements are working properly.

On the other hand, some researchers have discretely controlled the height differences. A narrowband microstrip array with grooved ground [15] employs grooves with a height difference of $0.25\lambda_0$ to cause normal reflection phase cancellation, where $\lambda_0$ is the free-space wavelength at 10 GHz. However, SCSR is obtained only near 10 GHz. In [16], a one-dimensional (1D) height multivariate optimization is proposed, which utilizes a series of narrowband microstrip antenna linear arrays with irregular heights and high isolation. It presents a remarkable 3 ~ 30 GHz ultrawideband property with large SCSRs. However, with the sharp height variations that destroy periodic electromagnetic environment, the optimization cannot guarantee broadband radiation and scanning. The radiation bandwidth is still less than 10%. It is important to find a way to simultaneously enhance radiation bandwidth, ensure the proper radiation of all elements, and maintain excellent low-scattering characteristics.

A stepped metal ground is integrated with a wideband dual-polarized dipole array in this communication. The proposed array is a 1D quasi-conformal array that exhibits ultra-wideband low-scattering characteristics as well as wideband radiation. The linking metal structure establishes vertical coupling between neighboring dipoles with height difference. This *x-y* plane dipole design allows the subarrays to synthesize scanning beams while maintaining a low standing wave ratio. The improved radiation achieves scanning from 0 to ± 45° under VSWR < 2.8 within 8 ~ 12 GHz (40%). The proposed 17 × 8 array prototype achieves a significant reduction in the monostatic SCS within 3.6 ~ 30 GHz with an averaged reduction of 19.4/18.9 dB, under normal *x/y* polarized incident waves respectively.

## II. Design Methodology

### A. Ground Scattering Analysis

Non-planar configurations are widely adopted in practice to deflect electromagnetic waves away from the incident direction. Consider a one-dimensional height distribution described by the undefined curve $h(x)$, where $x$ ranging from $[0, X_0]$ and $h(x)$ ranging from $[0, h_0]$, as shown in Fig. 1(a). The incident electric field and the monostatic backscattered electric field are denoted as $E_i$ and $E_s$. Under a geometric optics approximation, the reflection coefficient $\Gamma(f)$ can be expressed by integrating the fields from different parts of the ground:

$$\Gamma(f) = \frac{E_s}{E_i} = \frac{1}{E_i}\int_0^{X_0} e^{j[h_0 - h(x)]\frac{4\pi f}{c}} \frac{E_i dx}{X_0} \quad (1)$$

There is a spatial Fourier transform (SFT) relationship [17] between $\Gamma(f)$ and $h(x)$. Given the relationship between SCS-$f$ response and $\Gamma(f)$:

$$SCS(f) = \frac{4\pi A^2}{\lambda^2}|\Gamma(f)|^2 \quad (2)$$

where $A$ is the *xoy* aperture area, this shows how a non-planar metal ground imposes phase modulation on the backscattering field. The physical meaning is that, by considering only the *x-y* plane specular reflection, the spectral density of the reflected energy is reshaped by

This work was supported by the National Natural Science Foundation of China Projects under Grant U20A20165. (Corresponding author: Shi-Wei Qu)

The authors are with the School of Electronic Science and Engineering, University of Electronic Science and Technology of China, Chengdu 611731, China. (email: shiweiqu@uestc.edu.cn).

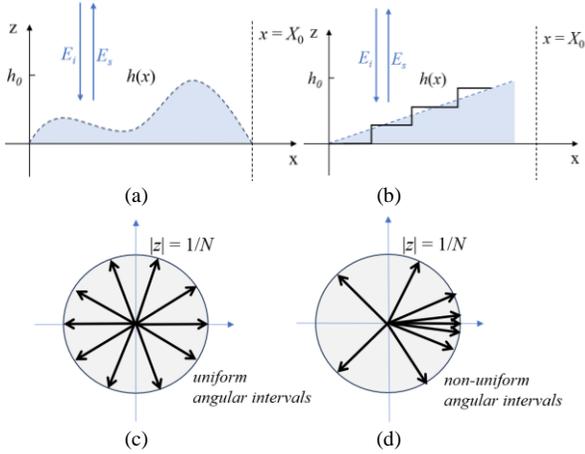

Fig. 1. (a) The one-dimensional height distribution $h(x)$ and (b) a linear discrete/continuous $h(x)$; The summation of reflected field vectors in complex plane with (c) linear and (d) nonlinear phase modulations.

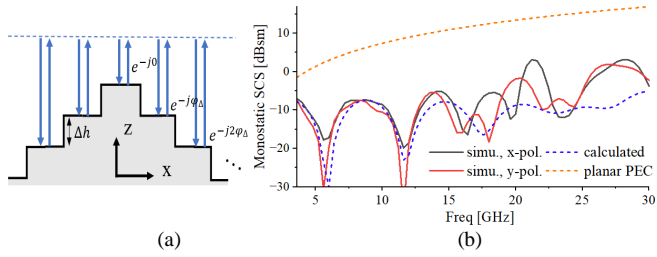

Fig. 2. (a) The proposed bilateral-linear stepped ground and (b) its computed and simulated monostatic SCSs under $N = 9$, $H = 0.1$.

the height differences $h(x)$. Assuming $h(x) = kx$ (where $k$ is the slope), the exponent in the integrand of (1) describes a linear phase modulation. The integration result for the continuous case is:

$$\Gamma_{cont}(f,k,X_0) = e^{j(h_0-kX_0/2)\frac{4\pi f}{c}} \mathrm{sinc}\left(\frac{\pi k f X_0}{c}\right) \quad (3)$$

which exhibits *sinc* function response, characteristic of a continuous spatial Fourier transform (CSFT). When $h(x) = h[n] = n\Delta h = n \cdot cH/f_c$ (for $n = 1, 2, ..., N$), the integration becomes a summation:

$$\Gamma(f,N,H) = \frac{E_s}{E_i} = \frac{1}{E_i}\sum_{n=1}^{N} e^{j[h_0-h[n]]\frac{4\pi f}{c}}\frac{E_i}{N} \quad (4)$$

$H$ is the electrical length with respect to $f_c$. The summation result for the discrete case is:

$$\Gamma_{disc}(f,N,H) = e^{j(N-1)\alpha/2}\frac{\sin(N\alpha/2)}{\sin(\alpha/2)}, \alpha = \frac{4\pi H f}{f_c} \quad (5)$$

which exhibits a $\sin(N\alpha)/\sin(\alpha)$ response (also known as the *Dirichlet kernel*), characteristic of a discrete spatial Fourier transform (DSFT).

A typical discretization evenly divides $h(x)$ into $N$ segments, each contributing equally to the amplitude of the backscattering field. Fig.1(c) and (d) illustrate the complex field vectors under linear and nonlinear phase modulations. To minimize backscattering over a wide bandwidth, the sum vector should remain close to the origin. A vector sequence with uniform phase intervals—as produced by a linear $h[n]$—will maintain their equal spacing at all frequencies. In contrast, due to an unequal amplitude distribution at different phase, the sum vectors of a nonlinear $h[n]$ generally deviate from the origin.

The proposed stepped ground, illustrated in Fig. 2(a), adopts a bilateral-linear discrete $h[n]$. Its monostatic SCSs can be calculated using (6) and compared with simulated results, as shown in Fig. 2(b).

$$SCS(f,N,H) = \frac{4\pi A^2}{\lambda^2}|\Gamma|^2 = \frac{4\pi A^2 f^2}{N^2 c^2}\left|\sum_{n=1}^{N-1} e^{-jn4\pi(h_0-h[n])f/f_c}\right|^2 \quad (6)$$

where $A = MD_yND_x$, $D_x$ and $D_y$ correspond to the element spacing in

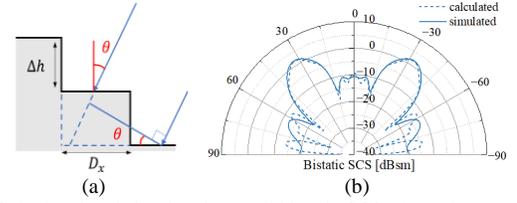

(a) (b)
Fig. 3. Calculated and simulated ground bistatic SCSs (*xoz*-plane) at 10 GHz.

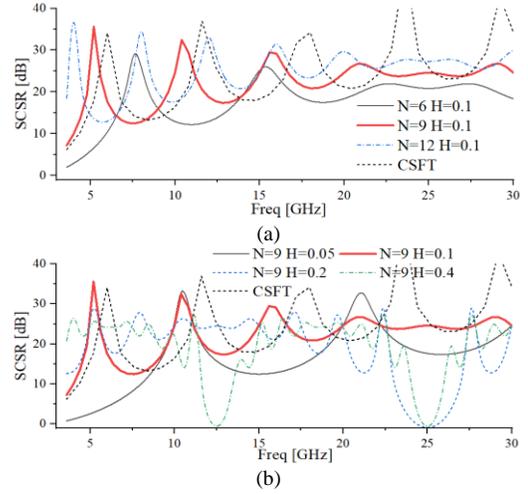

Fig. 4. Monostatic SCSRs (to a planar PEC) with different (a) $N$ and (b) $H$ under normal incidences.

$x$ and $y$-axis directions of the $N \times M$ array. Discrepancies between calculation and simulation arise because the calculation assumes that only specular reflection occurs from the *x-y* plane steps, neglecting edge/tip diffraction and other complex scattering components that are dependent on the incident polarization [18].

When electromagnetic waves are obliquely incident at an angle $\theta$ relative to the *z*-axis in the *xoz* plane, as shown in Fig. 3(a), $|\Gamma|$ and SCS can also be derived. The phase difference between two neighboring steps is:

$$\Delta\varphi = \frac{4\pi f}{c}\sin\theta\left(h'[n]\cot\theta - D_x\right) \quad (7)$$

where $h'[n] = h[n+1] - h[n]$ is the first-order difference. Similarly, the SCSs under oblique incidences can be derived:

$$SCS(\theta) = \frac{4\pi A^2 \cos^2\theta}{\lambda^2}|\Gamma|^2 = \frac{4\pi A^2 f^2 \cos^2\theta}{N^2 c^2}\left|\sum_{n=1}^{N-1} e^{-jn4\pi f(\sum_{m=1}^{n}\Delta\varphi_m)/c}\right|^2 \quad (8)$$

where $\Delta\varphi_m$ denotes the spatial phase difference between $n = m$ and $n = m+1$. The bistatic SCS of the proposed ground under normal incidence can be obtained as Fig. 3(b). Given the maximum height limitation, the proposed bilateral-linear $h[n]$ distribution produces lower bistatic SCSs near broadside than other possible distributions.

$N$ and $H$ are critical parameters influencing SCS-frequency response. A null of $|\Gamma|$ corresponds to a peak in SCSR, while the SCSR vanishes completely when $|\Gamma| = 1$. Fig. 4 reveals:

1) Parameter $N$ reflects the combined effect of the total height ($h_0$, *z*-axis direction boundary) and total length ($X_0$, *x*-axis direction boundary). As $N$ increases, the $|\Gamma|$ nulls become more densely distributed, accompanied by an overall increase in SCSR that particular enhances low-frequency performance;

2) An increase in $H$ essentially corresponds to a decrease in sampling rate. $H$ is closely related to the occurrence of $|\Gamma| = 1$ which aligns with the Nyquist sampling theorem; As $H$ increases, $|\Gamma| = 1$ appears at low frequencies; When $H$ is too small, $|\Gamma|$ nulls become relatively sparse and low-frequency SCSR is poor.

Excessive $H$ values prevent vertical coupling, degrading radiation performance; Conversely, insufficient $H$ provides inadequate slope angles on the stepped ground, limiting the SCSR angular range. After optimization, $H = 0.1$ in electrical length is selected. Although larger $N$ improves SCSR performance, $N = 9$ is chosen for verification due to size and cost constraints.

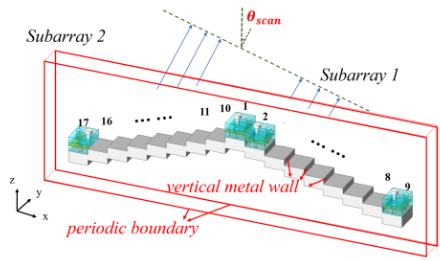

Fig. 5. Proposed 1D infinite unit.

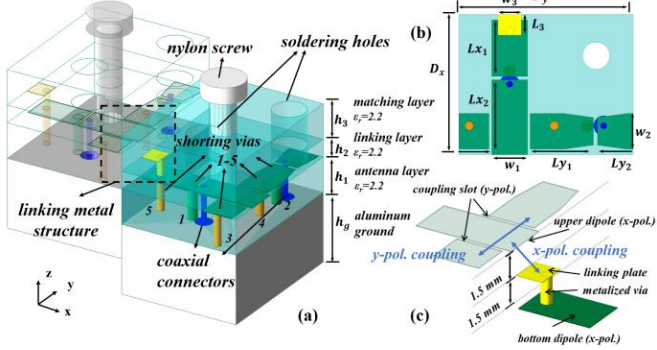

Fig. 6. Element topology (a) structure (b) design parameters (c) zoomed linking metal structure.

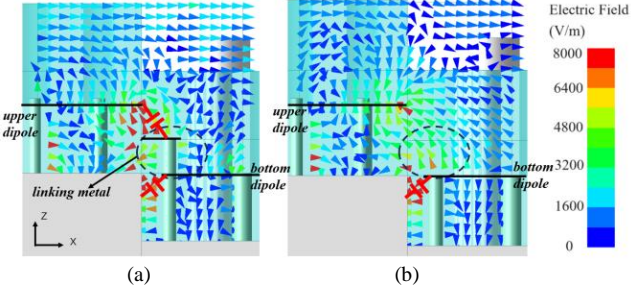

Fig. 7. Electric field distributions of the displacement currents (a) with (b) without linking metal structure.

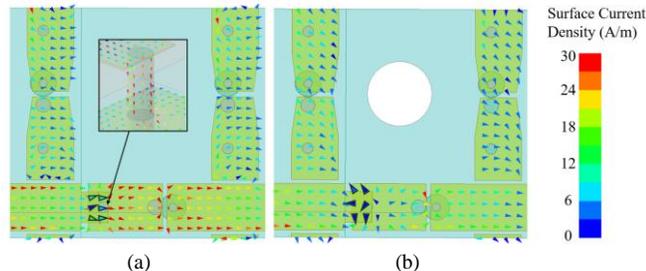

Fig. 8. Surface currents distributions on the dipoles, and on linking plate with bolded currents vectors (a) with (b) without the metalized via.

### B. Dipole Element

The proposed 17 × 8 prototype features a bilateral stepped ground structure along the $x$-axis and three-layer planar PCBs along the $y$-axis, fastened with nylon screws. Fig. 5 depicts a 1D infinite unit with periodic boundary conditions along the $y$-axis and 17 elements along the $x$-axis. The two subarrays are referred to as Subarray 1 (Elements 1-9) and Subarray 2 (Elements 10-17).

Previous height difference arrays typically adopt high isolated elements and irregular heights, which limit their radiation performance [15], [16]. The optical shadowing of metal walls also hinders the active participation of elements in scanning. The proposed dipole design enables simultaneous dual polarization and broadband radiation. Fig. 6 details the element topology, with dimensions of $10.2 \times 12.7 \times 8.9$ mm$^3$ ($0.408 \times 0.508 \times 0.356$ $\lambda_h^3$ at 12 GHz). As illustrated in Fig. 6(c), the $x$-polarized dipole exhibits vertical capacitive coupling through a parallel-plate configuration. Adjacent dipoles along the $y$-direction are indirectly coupled via an $x$-polarized dipole arm. The strength of this coupling varies inversely with the electrical length $H$ (normalized to $\lambda_c$). For $\Delta h = 1/10\lambda_c$ (3 mm), a linking layer with a thickness of 1.5 mm is added, which incorporates a metal sheet and a metalized via. Without this layer, dominant coupling between the bottom dipole end and ground plane would degrade the radiation of the $x$-polarized dipole, as illustrated in Fig. 7. In previous works such as [16], the radiation center frequency $f_c$ is set to a low value (4.6 GHz) to minimize $H$. In contrast, this work operates at a higher $f_c$ (10 GHz) within the SCSR band, resulting in a larger SCSR bandwidth below the radiation frequency ($0.36 f_c \sim f_c$), which is of significant practical importance.

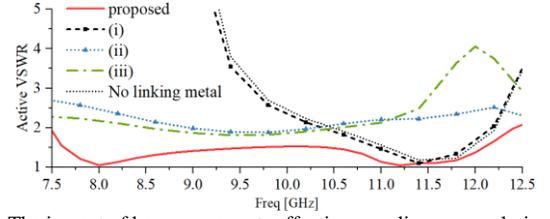

Fig. 9. The impact of key components affecting coupling on $x$-polarized active VSWRs at broadside (i) removing of the metallized via (ii) inward extension of the linking plate (iii) removing of Via 5.

TABLE I
KEY PARAMETERS AND HEIGHT DISTRIBUTION (UNIT: MM)

| $D_x$ | $D_y$ | $h_1, h_3$ | $h_2$ | $Lx_1$ | $Lx_2$ |
|---|---|---|---|---|---|
| 10.2 | 12.7 | 3 | 1.5 | 4.1 | 5.45 |
| $Ly_1$ | $Ly_2$ | $L_3$ | $w_1, w_2$ | $w_3$ | $h_g$ |
| 4.6 | 4.8 | 1.5 | 2.6 | 1.7 | 5.9 |
| $h[1]$ | $h[2], h[10]$ | $h[3], h[11]$ | … | $h[8], h[16]$ | $h[9], h[17]$ |
| 29.9 | 26.9 | 23.9 | … | 8.9 | 5.9 |

The linking metal structure couples the dipoles at different heights to establish consistent in-phase currents across the aperture [19]. This improves low-frequency VSWR by reducing reflections from the dipole arm ends, which would otherwise cause antiphase currents and undesired resonances. The strength of the displacement current and the phase consistency of the dipole currents are closely associated with three key components: (i) the metallized via, (ii) the linking plate, and (iii) Via 5. As illustrated in Fig. 8, the metallized via is critical for exciting a coherent radiation-mode current; its absence leads to weak radiation and also antiphase currents. Additionally, a larger size of the linking plate is not invariably beneficial. An inward extension of the linking plate would weaken the displacement current and bring about overall increase in VSWR. Via 5 also plays a key role by shifting the electric coupling from the ground to the upper dipole, strengthening the displacement current. The impact of these key components on the active VSWR is shown in Fig. 9.

Shorting vias 3, 4, and 5 in the antenna layer are meticulously tuned to improve impedance matching and shift the resonances to out-of-band frequencies [20].

The radiation fields of two subarrays are compensated to a common equiphase plane, which is parallel to the far-field wavefront as the beam is pointed to the angle $\theta_{scan}$. When scanning in one lateral direction, the opposing subarray must scan over a wider angular range to achieve this compensation. Taking Elements 5 and 13 (for Subarray 1 and 2) in the 1D infinite unit as representative elements, with both subarrays excited, the beam scanning within ± 45° in the E- and H-plane is achieved for dual-polarization over 7.6 ~ 12.1 GHz under an active VSWR < 2.8. This bandwidth is slightly wider than the standard X-band (8 ~ 12 GHz), providing margin for manufacturing tolerance. The simulated results are shown in Fig. 10.

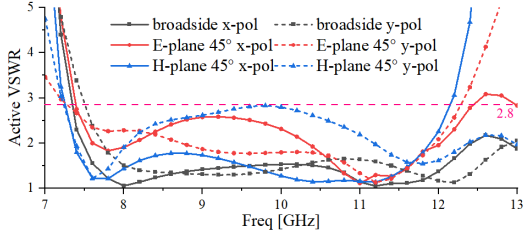

Fig. 10. Active VSWRs of the proposed 1D infinite unit (a) *x*-polarization (b) *y*-polarization.

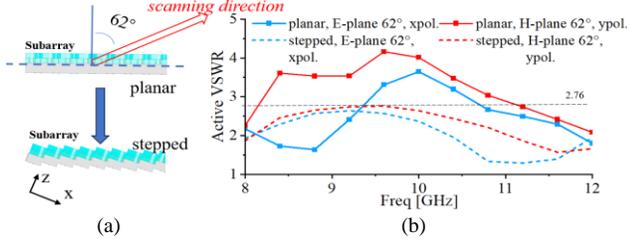

Fig. 11. Comparison of the (a) planar design and stepped design (b) active VSWRs of the Subarray 2 central element.

Fig. 11(a) illustrates the beam scanning of a planar subarray and the proposed stepped subarray. The dipoles in the planar subarray adopt conventional planar coupling, with similar impedance matching to that of the stepped element. From a slightly rotated perspective, it can be seen that each element in the proposed stepped subarray is tilted back towards the normal direction. To synthesize a +45° beam with the full array, the subarrays are required to scan beyond +62°. Conventional planar dipole elements exhibit rapid degradation in impedance matching under such conditions, as evidenced in Fig. 11(b). The *x-y* plane dipoles with linking metal structure make 62° scanning of the subarray in the *xoz* plane possible for both polarizations, while ensuring the proper radiation from all elements.

### C. Array Scattering Performance

A planar reference array with an identical aperture (173.4 × 114.3 mm$^2$) and similar impedance matching is designed and simulated for a valid scattering comparison. An equivalent continuous array with planar subarrays (tilted by 16.4°) in Fig. 11(a) is also designed. The element positions in this array are essentially identical to those of the proposed stepped array. Fig. 12 illustrates the progressive relationship among these three arrays. Whenever a planar array is transformed into a non-planar configuration, it can reduce the SCS within a certain angular range around the normal direction, albeit at the cost of radiation performance. Transforming a continuous array into a stepped array further enhances the scanning capability and also contributes to bistatic SCSR, as will be discussed later.

The simulated normal monostatic SCSs of the planar reference array and the proposed stepped array are shown in Fig. 13. Significant SCS reduction within an ultra-wideband range of 3.6 ~ 30 GHz is achieved. The averaged reductions in the normal monostatic SCS compared to the planar PEC plane are 22.6 dB and 22.8 dB, for *x*- and *y*-polarized incidences respectively. Compared to the reference array, the averaged reductions are 19.4 dB and 18.9 dB; the averaged in-band (8 ~ 12 GHz) reductions are 15.4 dB and 16.6 dB.

After loading antennas, the incident waves received by the ground is partially absorbed, which is the key to the relationship between the metal ground scattering properties in Section A and the array scattering results. The calculated SCSs in Fig. 13 are obtained using the following equation:

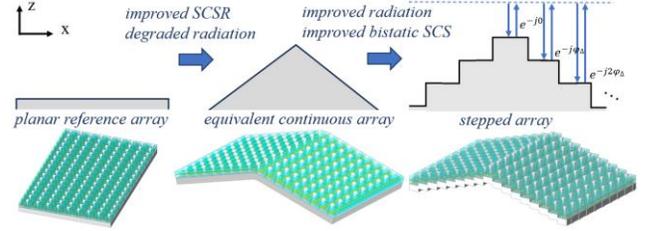

Fig. 12. The planar array, equivalent continuous array, and stepped array.

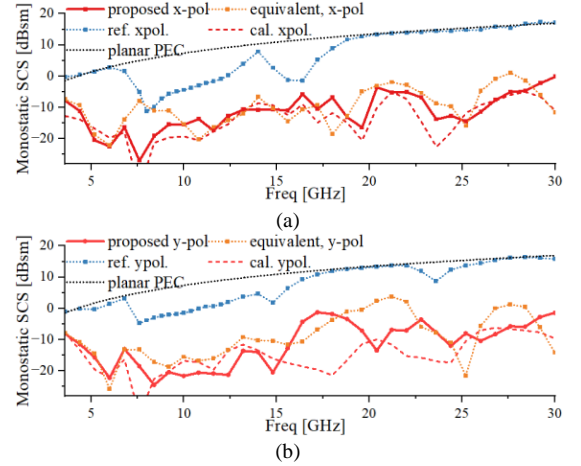

Fig. 13. Simulated and calculated monostatic SCSs of the three types of arrays under normal incidences (a) *x*-polarization (b) *y*-polarization.

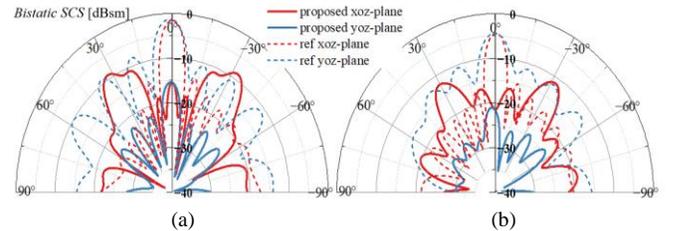

Fig. 14. The bistatic SCSs comparison between the proposed array and the reference array at 10 GHz (a) *x*-polarization (b) *y*-polarization.

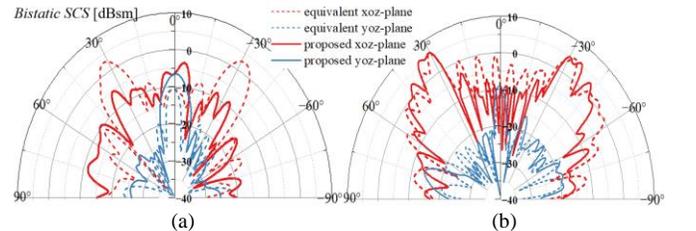

Fig. 15. The bistatic SCSs comparison between the proposed array and the equivalent continuous array at (a) 16.4 GHz (b) 24.4 GHz, *x*-polarization.

$$rEsca_{est} = rEsca_0 - \frac{\lambda}{2\eta P_0} \sum_{i=1}^{n} rEant_i \sum_{j=1}^{n} T_{ij} rEant_j \cdot E_{inc} \quad (9)$$

which approximate the ground without loaded antennas as an array with all antennas short circuited at the ground. Here $rEsca_0$ refers to the simulated ground scattering field; $\eta$ is free-space impedance, $n$ is total port number; $rEant$ refers to radiated field of antenna element under $P_0$ feeding power at the port; $E_{inc}$ refers to incident electric field. $T_{ij}$ is the element of $[T]$ defined by:

$$[T] = \frac{[\Gamma]}{[I] - [\Gamma][S]}\bigg|_{shorted} = \frac{[I]}{[I] + [S]} \quad (10)$$

$$[\Gamma] = diag(\Gamma_1, \Gamma_2, ..., \Gamma_n) \quad (11)$$

$[S]$ and $[I]$ denote the port scattering matrix and identity matrix; $\Gamma_i$ is the reflection coefficient between the load and the 50 Ω transmission line. The absorbed antenna mode field (the second term on the right side of (9)) is subtracted from the ground scattering field to obtain the scattering field of the proposed array. Under short-circuit conditions ($\Gamma_i = -1$), $[T]$ can be simplified. These calculations adhere to array scattering theory of antenna/structural mode scattering field [21]. The simulation-calculation discrepancies stem from the approximation process, primarily associated with the increased edge diffraction under *y*-polarized incidences.

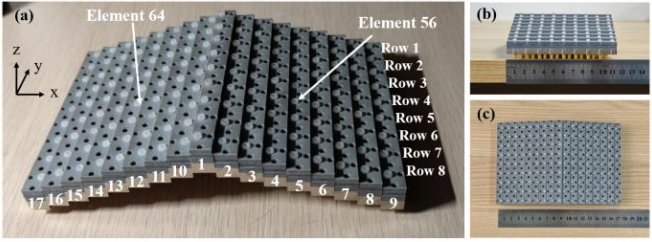

Fig. 15. The fabricated 17 × 8 prototype of the proposed array (a) 3D view (b) side view (c) top view.

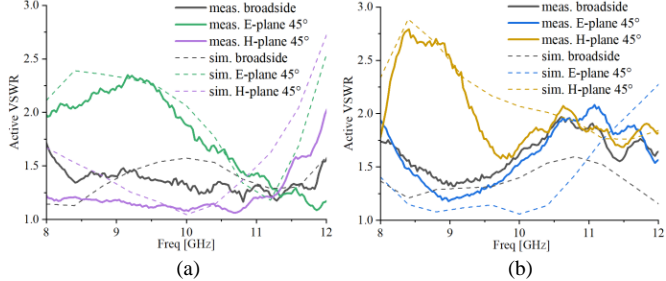

Fig. 16. Measured and simulated active VSWRs of the central elements (a) *x*-polarization (b) *y*-polarization.

The simulated bistatic SCSs at 10 GHz are illustrated in Fig. 14. The incident energy is scattered to two sides of the *xoz* plane. The reduced area of SCS possesses a groove shape, which means that the SCS in the entire *yoz* plane maintain low within ± 16° range in the *xoz* plane. This ensures the low-scattering characteristics within larger angular range in elevation plane, which is sufficient for most scenarios where the target is distant from the radar in azimuthal direction.

The monostatic SCSs of the equivalent continuous array are also presented in Fig. 13, demonstrating a strong similarity to those of the proposed array across the entire band. This consistency corroborates the approximate relationship between the CSFT and DSFT at a high sampling rate, which remains valid before and after antenna loading. Meanwhile, a more diffused bistatic SCS pattern within ±32° is achieved within 15 ~ 30 GHz. Two typical examples at 16.4 GHz and 24.4 GHz are shown in Fig. 15. At 16.4 GHz, although the normal SCS is slightly higher, the specular peaks at ± 32 ° disappear; At 24.4 GHz, an averaged bistatic SCS reduction of approximately 6 dB is observed within ±25°. The bistatic SCS becomes more diffused primarily due to the increased incoherent field components from surface roughness, as the stepped geometry introduces significant phase perturbations and violates specular reflection conditions. This effect is particularly pronounced at higher frequencies where structural details become electromagnetically relevant [22].

## III. EXPERIMENTAL RESULTS

A prototype of the proposed 17 × 8 array is fabricated, as illustrated in Fig. 15, with total sizes of 173.4 × 114.3 × 38.8 mm³. Passive reflection and coupling S-parameters are measured with a vector network analyzer (ZNA43). Elements 56 and 64 are designated as the representative central elements for Subarray 1 and Subarray 2. Active VSWRs of the two central elements are derived through synthesis of the passive S-parameters. The measured active VSWRs are compared with the simulated results, as shown in Fig. 16. The measured and simulated curves reveal consistent trends under the condition of VSWR < 2.8. Discrepancies can be attributed to the imperfect fastening tolerances and measurement inconsistencies.

Radiation measurements are conducted in an anechoic chamber. The active element patterns (AEPs) are measured (other ports terminated with 50 Ω loads) and the total patterns are synthesized using simulation feeding phases. The measured and simulated patterns of 8 GHz and 12 GHz are shown in Fig. 17 and Fig. 18, for *x*- and *y*-polarization respectively. The measured and simulated curves are generally in good agreement. The measured cr-pol levels reach at least -25 dB at broadside and -15 dB when scanning.

The scattering measurement setup includes T/R horns, signal generators, a receiver, a compact reflector, and a turntable (for oblique incidence measurement). The measured monostatic SCS-*f* results are shown in Fig. 19. The measured and simulated results are in good agreement. The measured results exhibit oscillatory deviations from simulation across the band while maintaining consistent trend. The measured and simulated monostatic SCSs under obliquely incidences in the *xoz* plane at 10 GHz are shown in Fig. 20, with an acceptable agreement. Similar SCS can be seen at 0°, and the measured SCSs are generally lower in the low-SCS region within ± 8°.

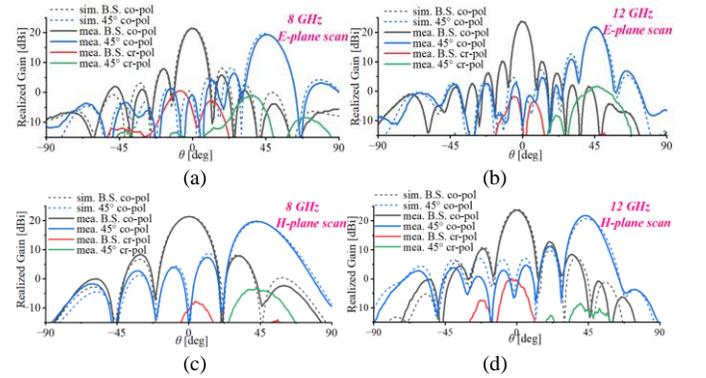

Fig. 17. Radiation patterns of *x*-polarization (a) 8 GHz E-plane (b) 12 GHz E-plane (c) 8 GHz H-plane (d) 12 GHz H-plane (B.S.: broadside).

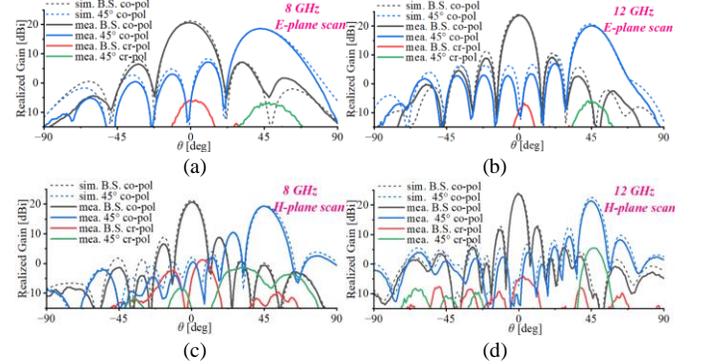

Fig. 18. Radiation patterns of *y*-polarization (a) 8 GHz E-plane (b) 12 GHz E-plane (c) 8 GHz H-plane (d) 12 GHz H-plane.

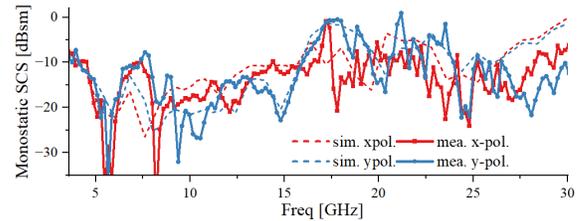

Fig. 19. Simulated and measured monostatic SCSs under normal incidences.

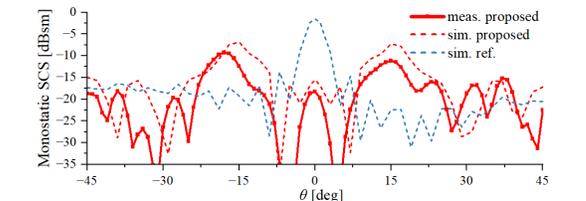

Fig. 20. Simulated and measured monostatic SCSs varying with the incident angles in the *xoz* plane, *x*-polarization incidences.

TABLE II
COMPARISON BETWEEN RECENT LOW-SCATTERING ARRAYS AND THIS WORK

| reference & technique | SCSR band | averaged SCSR | SCSR incident angular range | radiation band | scanning angular range | AE & RE[3] | radiation pol. | SCSR pol. |
|---|---|---|---|---|---|---|---|---|
| [7], feed optimization | 8.5-12 GHz (34%) | 6.4 dB | ±12° (*xoz*) | 8-12 GHz (40%) | ±60° (E) | N.A. | single | single |
| [10], chessboard array | 6-18 GHz (100%) | 10 dB | ±45° (*xoz/yoz*) | 8-12 GHz (40%) | ±60° (E/H) | N.A. | single | single |
| [15], height difference | 7.8-13 GHz (50%) | 9 dB[1] | ±8°/±45° (*xoz/yoz*) | 10-10.6 GHz (5.8%) | ±60°/45° (E/H) | N.A. | single | dual |
| [16], height difference | 3-30 GHz (164%) | 15 dB | ±4° (*xoz*)[1] | 4.4-4.8 GHz (8.7%) | ±60° (E) | 80%/N.A. | single | dual |
| This Work | 3.6-30 GHz (157%) | 18.9 dB | ±8°/±45° (*xoz/yoz*) | 8-12 GHz (40%) | ±45° (E/H)[2] | 85%/95% | dual | dual |

[1] Not given, and estimated by the figure. [2] For the subarrays, the *xoz*-plane scanning angle reaches 62°.
[3] AE & RE: Aperture efficiency and radiation efficiency.

Table II provides a comparison of this work against the recent studies on low-scattering phased arrays. Among designs employing different SCSR techniques, the height difference approach offers the distinct advantages of ultra-wideband reduction and polarization insensitivity to incident waves. Compared to previous works utilizing height differences, this work demonstrates enhanced radiation bandwidth and dual-polarized operation while maintaining scattering performance. Compared to [16], a larger SCSR angular range is achieved with a less vertical space occupation; SCSR are achieved in a frequency band lower than that of the operating radiation band.

## IV. CONCLUSION

This communication presents a wideband, low-scattering, dual-polarized phased array, featuring a stepped quasi-conformal ground and coordinated wideband dipoles with vertical coupling for improved radiation performance. Compared to prior studies on low-scattering arrays, this design leverages height differences to optimize scattering performance while mitigating their typical radiation drawbacks. The deployable antenna types of height difference arrays are expanded to dual-polarized element with inter-element coupling, adapting to more complex scenarios; the proposed array also exhibits both in-band and out-of-band SCS reduction under dual-polarized incidences, with the potential for application in multifunctional stealth systems.